\documentclass[12pt,twoside]{article} 
\usepackage{graphicx}

\date{} 

\begin{document} 

%\centerline{\bf Journal's Title, Vol. x, 200x, no. xx, xxx - xxx} 

\centerline{} 

\centerline{} 

\centerline {\Large{\bf Large Frequency Range of Photonic Band Gaps on Porous}} 

\centerline{} 

\centerline{\Large{\bf Silicon Heterostructures for Infrared Applications}} 

\centerline{} 

\centerline{\bf {J. Manzanares-Martinez and P. Castro-Garay}} 

\centerline{} 

\centerline{Centro de Investigacion en Fisica de la Universidad de Sonora,} 

\centerline{Apartado Postal 5-088, Hermosillo, Sonora 83190, Mexico.}  

\centerline{} 

\centerline{\bf {R. Archuleta-Garcia and D. Moctezuma-Enriquez}} 

\centerline{} 

\centerline{Programa de Posgrado en Ciencias (Fisica) de la Universidad de Sonora,} 

\centerline{Calle Rosales y Boulevard Luis Encinas, Hermosillo, Sonora, Mexico.} 

\newtheorem{Theorem}{\quad Theorem}[section] 

\newtheorem{Definition}[Theorem]{\quad Definition} 

\newtheorem{Corollary}[Theorem]{\quad Corollary} 

\newtheorem{Lemma}[Theorem]{\quad Lemma} 

\newtheorem{Example}[Theorem]{\quad Example} 

\begin{abstract} 
In this work we  show theoretically that it is possible to design a large 
band gap in the infrared range  using a one-dimensional Photonic Crystal heterostructure
 made of porous silicon. 
Stacking together multiple photonic crystal substructures
of the same contrast index , but of different lattice  periods, it  is possible to broad 
the narrow  forbidden band gap that can be reached by the low contrast  index 
of the porous silicon multilayers.
The main idea in this work is that we  can construct a Giant Photonic Band 
Gap -as large as desired- by combining a tandem of photonic crystals substructures
by using a simple analytical rule to determine the period of each substructure.
\end{abstract} 


\section{Introduction} 

In recent years, there has been much interest in the fabrication and study of Photonic Crystals (PCs)
\cite{Yablo,John} which are periodic dielectric composites with a periodicity 
matching the wavelength of electromagnetic
waves in the visible or  infrared range. \cite{infrared1,infrared2,infrared3} The spatial modulation 
of the refractive index leads to Photonic Band Gaps (PBG) where the electromagnetic fields 
cannot propagate due to the destructive interference. The absence of electromagnetic 
states in the PBG  allow the possibility that the PCs can be perfect mirrors 
for light impinging from any direction or polarization. \cite{Winn}
These  PC-based mirrors have low losses as 
compared with old metallic mirrors,
specially in the infrared range. \cite{Zhai} 

Theoretical photonic band structure calculations predict complete band gaps  for a number 
of periodic dielectric lattices in two (2D) or three (3D) dimensions. \cite{Joannopoulos} 
Nevertheless, to fabricate  3D PCs with a band gap in the infrared is a  difficult and
time consuming task. \cite{Lai} Additionally, even with the most recent experimental techniques 
it has only been possible to fabricate 3D structures with a low variation in the refractive 
index, typically $n<2.0$.\cite{Zhai}  This low index contrast attained in  3D PCs
is a limiting factor in some applications in telecoms or near infrared range
where  a large range of band gap is required. \cite{Dhillon,Zlatanovic} 
To obtain wider PBG it is desirable to increase the refractive index,
different  authors  have reported that the width of the band gap is 
a function of the contrast index, 
the larger the better. \cite{Yablo_2000,Wijnhoven_1998,Chan_2005}

 Recently   several improvements in the 
fabrication of one-dimensional (1D) photonic crystals or multilayers
 based in  porous silicon have been reported. \cite{Agarwal1,Palestino} 
Porous silicon fabrication techniques are 
flexible, cheap and  feasible to grow with a high number of periods in 
the infrared range.\cite{Agarwal2,Wang} Porous silicon multilayers have 
 a low index contrast, however. \cite{Zhai} Nevertheless, it has been reported that
the band gap can be enlarged by using heterostructures. \cite{Zi}
A heterostructure is a mirror composed by a tandem of  PC substructures or submirrors 
and the idea is to  add the band gaps of each submirror. \cite{Istrate} 
Some of the present authors
have previously  reported the design of a heterostructure 
mirror in the complete visible.\cite{Archuleta_2010} 
In this work, we present the design of a photonic heterostructure mirror 
in the near infrared range
($0.65$ $\mu m$-$3.0$ $\mu m$). Our theoretical approach shows that it is possible to
obtain a near infrared mirror with fewer layers than those needed in an already
 published work, 
\cite{Agarwal1} where the heterostructure mirror was designed  following 
an experimental rule. \cite{Agarwal2} 

\begin{figure}
\includegraphics[scale=0.75,clip]{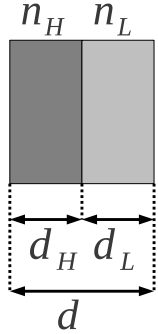}
\centering
\caption{\label{Fig_1} Unit cell composed of two materials with indices $n_H$ and $n_L$.
The width of the unit cell is $d=d_H+d_L$. }
\end{figure}

\section{Theory}

Let us first consider a PC made by the infinite periodic repetition of the unit cell
 illustrated in Fig. 1. 
The unit cell is composed by two materials, one is of  
high ($n_H$) and the other of low ($n_L$) refractive index.  
The  width of each layer is $d_H$ and $d_L$, respectively. 
The unit cell have a width of $d=d_H+d_L$. 
The  Photonic Band Structure is obtained using an 
analytical formula derived by Yeh {\it et al.} 
\cite{Yeh}

\begin{equation}
 cos(kd)=
\cos 
\left( 
\frac{n_H \omega d_H}{c}
\right)
\cos 
\left( 
\frac{n_L \omega d_L}{c}
\right) -
\frac{1}{2}
\left(
\frac{n_H}{n_L}+\frac{n_L}{n_H}
\right)
\sin 
\left( 
\frac{n_H \omega d_H}{c}
\right)
\sin 
\left( 
\frac{n_L \omega d_L}{c}
\right)
\end{equation}

$k$, $\omega$ and $c$ are the wavevector, frequency and speed of light, respectively.  
It is convenient to write this equation in the form

\begin{equation}
 cos(kd)=
\cos 
\left( 
2 \pi n_H f_H \Omega
\right)
\cos 
\left( 
2 \pi n_L f_L \Omega
\right) -
\frac{1}{2}
\left(
\frac{n_H}{n_L}+\frac{n_L}{n_H}
\right)
\sin 
\left( 
2 \pi n_H f_H \Omega
\right)
\sin 
\left( 
2 \pi n_L f_L \Omega
\right),
\end{equation}

where we have introduced the  filling fraction
as the space occuped by each material in the unit cell,
 $f_H=d_H/d$ and $f_L=d_L/d$. The reduced frequency is

\begin{equation}
 \Omega = \frac{d}{\lambda}
\end{equation}

where we have considered that $c=\lambda/\tau$ and $\omega=2\pi/\tau$.

\begin{figure}
\includegraphics[scale=0.65,clip]{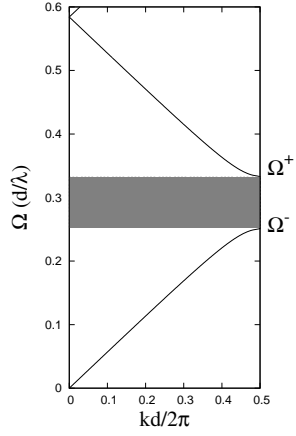}
\centering
\caption{Photonic Band Structure of a crystal with refractive
indices  $n_H=2.2$ and $n_L=1.4$. The filling fractions are $f_H=0.39$ and $f_L=0.61$.
The band gap has an upper and lower reduced 
frequency limit of $\Omega^+=0.33$ and $\Omega^-=0.25$, 
respectively}
\end{figure}

We consider a unit cell composed by two high and low typical
porous silicon  refractive  indices values, $n_H=2.2$ and  $n_L=1.4$. 
The filling fractions that optimizes the width of the first 
photonic band gap are calculated by using the 
quarter wave condition $n_H d_H=n_L d_L=\lambda/4$, \cite{Sozuer} 
obtaining $f_H=0.39$ and $f_L=0.61$, respectively.
In Fig. 2 we present the photonic band structure. The gray zone 
is the first forbidden band gap which is limited by an upper and lower reduced frequency 
with values
 $\Omega^+=0.33$ and $\Omega^-=0.25$, respectively. These limits can be related to the
 wavelength by using eq. (3) 

\begin{equation}
 \lambda^+(d) = \frac{d}{\Omega^-}
\end{equation}

\begin{equation}
 \lambda^-(d) = \frac{d}{\Omega^+}
\end{equation}

\begin{figure}
\centering
\includegraphics[scale=0.65,clip]{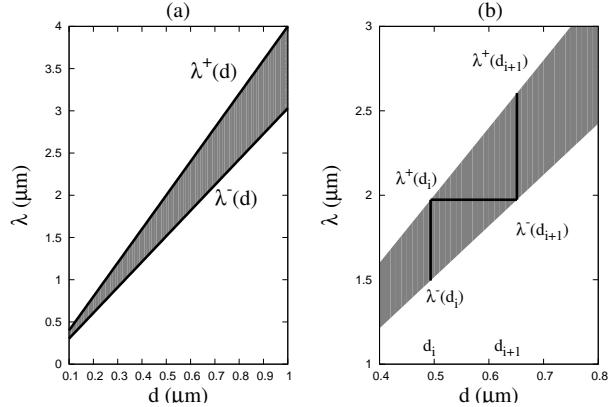}
\centering
\caption{Variation of the band gap (gray zone) as a function of the period.
In panel (a) we present the variation of the limits $\lambda^+(d)$ and $\lambda^-(d)$ 
as a function of $d$. Panel (b) illustrates the increase of the band gaps
taking two PC-lattices of periods $d_i$ and $d_{i+1}$. }
\end{figure}

In Fig. 3(a) we plot in the gray zone the enlargement of the  band gap
as a function of the period, $d$.
The upper and lower limits of the band gap are
$\lambda^+(d)$ and $\lambda^-(d)$, respectively. 
To design a heterostructure with a giant band gap, we stack
together PC lattices or submirrors of different periods
such that the band gaps can be added. In panel (b) we 
illustrate the addition of the band gaps of two lattices of periods
$d_i$ and $d_{i+1}$. The submirror of period $d_i$ ($d_{i+1}$)
has upper and lower limits of $\lambda^+_i$ ($\lambda^+_{i+1}$) and
$\lambda^-_i$ ($\lambda^-_{i+1}$), respectively.
The purpose is to choose the periods so that 
the upper limit of one submirror is equal to the lower limit of the next submirror, 
as follows

\begin{equation}
 \lambda^+(d_i)=\lambda^-(d_{i+1})
\end{equation}

Using eqs. (4), (5) and (7) this relation can be written as a rule for the periods
in the form

\begin{equation}
 d_{i+1}=\frac{\Omega^+}{\Omega^-}d_i
\end{equation}

\begin{figure}
\includegraphics[scale=0.6,clip]{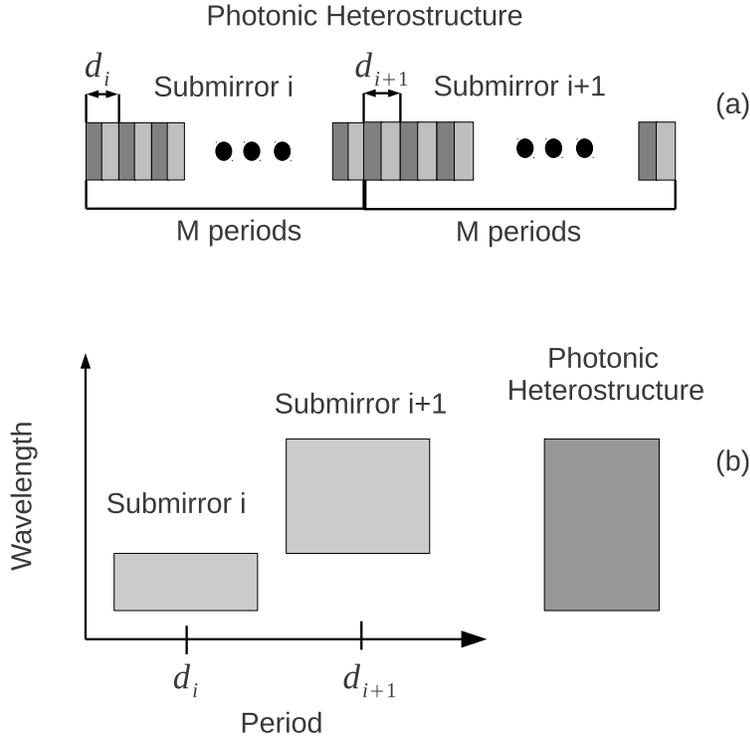}
\centering
\caption{Schematic of the heterostructure composed by the addition of two submirrors
of periods $d_i$ and $d_{i+1}$. Panel (a) shows a heterostructure composed by 
 the union of two submirros. In panel (b)
we present a representation of the addition of the band gaps. }
\end{figure}

Fig. 4 presents in  panel (a) a PC heterostructure formed by two submirros with a lattice period of 
$d_i$ and $d_i$, respectively. Each submirror has $M$ unit cell periods. 
In panel (b) we illustrate the addition of the photonic band gap of each submirror to
make a wider heterostructure band gap.

\section{Numerical results}

In order to present our method to design a photonic crystal heterostructure
with an enlarged band gap we consider the case of 
a mirror for  the whole near infrared range, specifically with a low and high frequency 
limits of $\lambda_{high}=0.3$ $\mu m$  and $\lambda_{low}=0.65$ $\mu m$, respectively. 
To select the thickness of the first submirror $d_1$,  we consider 
that its lower frequency limit $\lambda^-(d_1)$ is equal to the low frequency 
limit of the heterostructure mirror, $\lambda_{low}$. This condition can be written in the form

\begin{equation}
 \lambda(d_1)=\lambda_{low}
\end{equation}

The first submirror period is obtained using the relation

\begin{equation}
 d_1 = \frac{\lambda_{low}}{\Omega^+}
\end{equation}

To determine the following submirror periods $d_i$ ($i>1$) we use
eq. (7). For each submirror, the high and low
frequency limits are calculated using eqs. (4) and (5).
The idea is to calculate the period and the
frequency limits for $N$ submirrors until 
a condition is reached 
where the high frequency limit of the $N$ submirror is greater than the high 
frequency limit of the desired heterostructure mirror $\lambda_{high}$,

\begin{equation}
 \lambda^+(d_N) > \lambda_{high}
\end{equation}

\begin{figure}
\includegraphics[scale=0.8,clip]{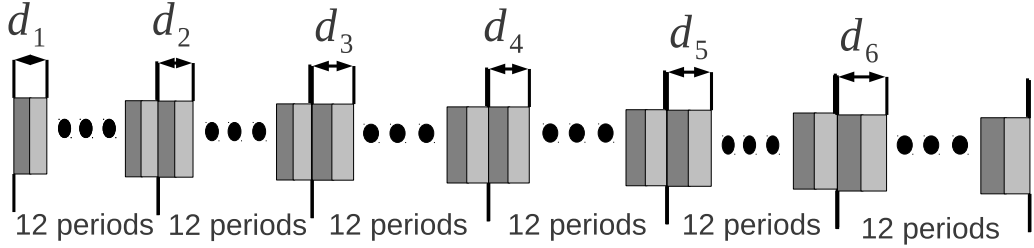}
\centering
\caption{Schematic of the Photonic Crystal heterostructure composed by 6 submirrors.
Each submirror have 12 unit cell periods.}
\end{figure}

For the case of the porous silicon PC calculated in the previous section,
we have found that with 6 submirrors can cover the whole near infrared range.
In table 1 we present the periods and band gaps limits calculated  for each submirror.
Fig. 5 shows a schematic of the photonic crystal heterostructure composed by six
submirros, where each submirror posses 12 unit cells periods. The complete 
heterostructure has 72 unit cells. In Fig. 6 we present the 
calculated light reflection of the photonic heterostructure. The simulation program 
used is based on the transfer matrix method \cite{Yeh}.
We observe that  total reflection occurs in the range $0.65 \mu m$-$3.43 \mu m$.

\begin{table}[h]
\renewcommand{\arraystretch}{1.3}
\caption{Table of the periods ($d_i$) and the upper and lower
band gap limits ($\lambda^+_i$ and $\lambda^-_i$) of each submirror}
\vskip0.2in
\begin{center}
\small \begin{tabular}{|c|c|c|c|}\hline
submirror & $d_i$ ($\mu$m) & $\lambda_i^+$ ($\mu$m)& $\lambda_i^-$ ($\mu$m) \\ \hline
1 & 0.21 & 0.85  & 0.65 \\ \hline
2 & 0.28 & 1.13  & 0.85 \\ \hline
3 & 0.37 & 1.49  & 1.13 \\ \hline
4 & 0.49 & 1.97  & 1.49 \\ \hline
5 & 0.65 & 2.60  & 1.97 \\ \hline
6 & 0.86 & 3.43  & 2.60 \\ \hline
  \end{tabular}
\end{center}
\label{tab1}
\end{table}

 \begin{figure}
\includegraphics[scale=0.9,clip]{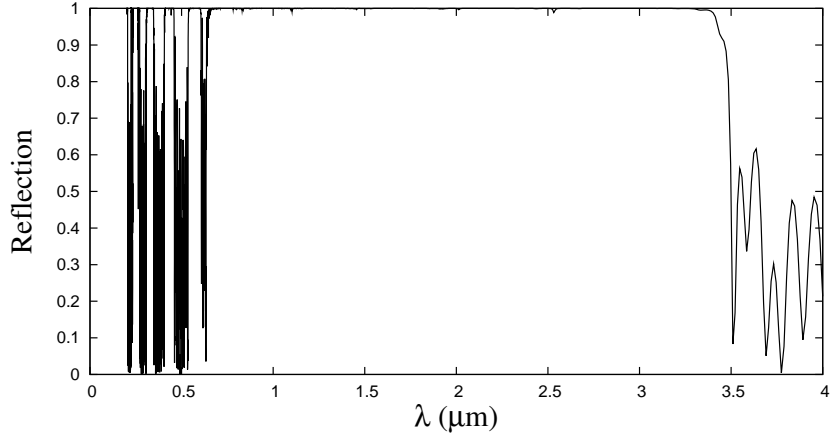}
\centering
\caption{Light reflection from a Photonic Heterostructure mirror composed by six
submirrors. The period of each submirror is presented in Table 1.}
\end{figure}

\section{Discussion}

We have demonstrated that a  
porous silicon  heterostructure 
reflector with low refractive contrast index   can be designed by using a simple 
rule to determine the periods. The near infrared reflector consists of 6 
submirrors with 12 unit cells periods each. The complete structure has 72 unit cells
and reflects light in the  infrared range of  $0.65 \mu m$-$3.43 \mu m$.
In comparison, in Ref. \cite{Agarwal1} it has been reported the fabrication of a heterostructure
 mirror for the near infrared range which is 
composed by 45 submirrors of 4 periods for each submirror, obtaining a total of 180 unit cells.
The heterostructure that we have designed in section III has  $40\%$ fewer layers as compared to
Ref. \cite{Agarwal1}. On the other hand, our formula to determine the submirror periods
is based on  a clear physical insight related to the addition of the photonic band gaps,
while it is not clear the reason why the empirical  
rule proposed Refs. \cite{Agarwal1,Agarwal2} should work.

\section{Conclusion}
 We have presented a simple strategy to design a 
heterostructure mirror composed by a stack of several submirrors
with the same refractive indices $n_H$ and $n_L$ and different period thicknesses $d_i$.
Stacking together different substructures the band gap of the heterostructure
is the union of the band gap of the submirrors. For this reason, each submirror
has been chosen so that the band gaps can be added.
 We have presented the design of a mirror for the near infrared range. However, 
our method  can be used to design a heterostructure with a 
band gap as large  as wide
which can work in any frequency range as far as the refractive indices of the
composite materials remains valid. 
 
The proposed porous silicon based photonic heterostructure has been designed 
using parameters of refractive indices and periods 
that are  compatible with the well established porous silicon
technology. These structures 
have an excellent potential for integration with optoelectronic
technology. This is relevant because one of the goals of silicon based 
fabrication techniques is to integrate
infrared or optical components such as modulators, photodetectors or 
sensors in porous silicon ships to fabricate new photonic devices.

{\bf ACKNOWLEDGEMENTS.} 
PCG thanks the Optical Academy of the Department for Physics
Research of the Sonora University for a temporary contract as Research
Associate. RAG acknowledges a grant provided by the CONACYT project CB-2008-104151.
We thank the head of our Department for Physics Research, Prof. Julio Saucedo
Morales for reading the manuscript.

{\bf Received: Month xx, 200x}

\end{document}